\documentclass[journal]{IEEEtran}
\usepackage{ifpdf}
\usepackage{url}
\usepackage{color}
\usepackage{soul}
\usepackage[normalem]{ulem}
\usepackage{units}
\ifpdf
\else
\fi

\usepackage{cite}
\ifCLASSINFOpdf
  \usepackage[pdftex]{graphicx}
  \graphicspath{{../pdf/}{../jpeg/}}
  \DeclareGraphicsExtensions{.pdf,.jpeg,.png}
\else
  \usepackage[dvips]{graphicx}
  \graphicspath{{../eps/}}
  \DeclareGraphicsExtensions{.eps}
\fi

\usepackage{amsmath}
\usepackage{amsfonts}
\usepackage{amssymb}
\usepackage{algorithm}
\usepackage{algpseudocode}
\algtext*{EndWhile}
\algtext*{EndIf}
\algtext*{EndFor}

\usepackage{array}

\ifCLASSOPTIONcompsoc
 \usepackage[caption=false,font=normalsize,labelfont=sf,textfont=sf]{subfig}
\else
 \usepackage[caption=false,font=footnotesize]{subfig}
\fi

\usepackage{stfloats}

\usepackage{hyperref}

\long\def\comment#1{}

\DeclareMathOperator*{\argmax}{arg\,max}
\DeclareMathOperator*{\argmin}{arg\,min}

\newfont{\bbb}{msbm10 scaled 700}

\newfont{\bb}{msbm10 scaled 1100}

\newcommand{\av}{{\bf a}}
\newcommand{\bv}{{\bf b}}

\newcommand{\dv}{{\bf d}}

\newcommand{\gv}{{\bf g}}

\newcommand{\nv}{{\bf n}}

\newcommand{\uv}{{\bf u}}
\newcommand{\wv}{{\bf w}}

\newcommand{\xv}{{\bf x}}
\newcommand{\yv}{{\bf y}}

\newcommand{\Rm}{{\bf R}}

\newcommand{\Wm}{{\bf W}}

\begin{document}

\title{Antenna Selection in Switch-Based MIMO Arrays via DOA threshold region Approximation}

\author{Hui~Chen,~\IEEEmembership{Member,~IEEE}, Tarig~Ballal,~\IEEEmembership{Member,~IEEE}, 
Mohammed~E.~Eltayeb,~\IEEEmembership{Member,~IEEE},  and~Tareq~Y.~Al-Naffouri,~\IEEEmembership{Senior Member,~IEEE}
\thanks{Hui~Chen, Tarig~Ballal, and  Tareq~Y.~Al-Naffouri are with the Division of Computer, Electrical and Mathematical Science \& Engineering, King Abdullah University of Science and Technology (KAUST), Thuwal, 23955-6900, KSA. e-mail: (\{hui.chen; tarig.ahmed; tareq.alnaffouri\}@kaust.edu.sa). Mohammed E. Eltayeb is with California State University, Sacramento, USA. (Email: mohammed.eltayeb@csus.edu)}}

\maketitle
\begin{abstract}
Direction-of-arrival (DOA) information is vital for multiple-input-multiple-output (MIMO) systems to complete localization and beamforming tasks. Switched antenna arrays have recently emerged as an effective solution to reduce the cost and power consumption of MIMO systems. {Switch-based array architectures connect a limited number of radio frequency chains to a subset of the antenna elements forming a subarray. This paper addresses the problem of antenna selection to optimize DOA estimation performance. We first perform a subarray layout alignment process to remove subarrays with identical beampatterns and create a unique subarray set. By using this set, and based on a DOA threshold region performance approximation,} we propose two antenna selection algorithms; a greedy algorithm and a deep-learning-based algorithm. The performance of the proposed algorithms is evaluated numerically. The results show a significant performance improvement over selected benchmark approaches {in terms of DOA estimation in the threshold region and computational complexity}. 

\end{abstract}


\IEEEpeerreviewmaketitle
\section{Introduction}
{Direction-of-arrival (DOA) is an important phase in localization,} which assists a variety of applications, such as location-aware communications~\cite{talvitie2019positioning}, vehicular networks~\cite{5-vehicle}, virtual reality~\cite{ge2017multipath}, etc.
Compared with the time-of-arrival/time-difference-of-arrival (TOA/TDOA) where precise clock synchronization is needed, DOA is preferred in localization~\cite{3-airwriting}.
{Furthermore, orientation estimation becomes possible with DOA information at the user side in high-frequency communication systems (e.g., mm-Waves and THz systems)~\cite{chen2021tutorial}. Practical DOA estimation algorithms, such as MUSIC~\cite{doa_music} and ESPRIT~\cite{doa_esprit}, are widely used to solve this problem.}



{The DOA estimation algorithms are usually used in digital arrays; however, assigning a radio-frequency chain (RFC) to each antenna in high-frequency communication systems is impractical due to the large array size.} {Hybrid analog-digital architectures are proposed in multiple-input-multiple-output (MIMO) systems with the implementation of phase shifters (PSs), switches, or lenses~\cite{8-heath2016overview}. The PS-based structures are usually considered in MIMO systems; nevertheless, the switch-based architectures have drawn researchers' attention due to its reduced complexity and low power consumption~\cite{mendez2016hybrid}, which is crucial for the mobile device.} For a switch-based array, a small number of RFCs are associated with a subset of antennas. Considering that the geometry of the selected subarray affects DOA estimation, a performance-optimizing criterion needs to be applied to the antenna selection process. 


To characterize DOA estimation performance for a specific array layout and SNR, the \emph{mean squared error} (MSE) is usually adopted. {The MSE of an unbiased (DOA) estimator is bounded by the Cram\'er-Rao lower bound (CRLB), which can be used as an antenna selection criterion to optimize DOA estimation performance by choosing the subarray that minimizes the CRLB~\cite{only_crlb_01}.} However, when the SNR falls below a certain threshold, the large sidelobes {of the array beampattern} contribute to an abrupt increase in the MSE of a DOA estimator. {This phenomenon is known as the} \emph{threshold effect}, and it renders the CRLB infeasible in the low SNR regime~\cite{threshold_tsp, threshold_expression}. {In other words, selecting antennas based on CRLB does not guarantee a good DOA estimation performance.}

{The impact of the threshold effect on antenna selection can be alleviated by adding constraints on the peak sidelobe level (PSL) and leveraging convex optimization tools to minimize the CRLB~\cite{wang2014adaptive,gupta_subarray}.} In spite of the great performance, the application of a fixed threshold level across the whole DOA range does not guarantee optimality since the threshold value is DOA-dependent.
Deep-learning-based techniques can be employed to reduce the computational complexity of the selection process, as in~\cite{huang2018deep,cao2020complex}. Unfortunately, in these works, the threshold region effect is not considered. This necessitates the design of new antenna selection algorithms that are applicable to all incident DOAs, especially in the low SNR regime.

In this paper, we propose two antenna selection algorithms, namely, a greedy algorithm and a deep learning (DL)-based algorithm based on threshold region approximation (TRA), for switch-based MIMO architectures to avoid the above-highlighted issues and achieve better DOA estimation performance. In Section~\ref{problem_statement}, we introduce the antenna-selection problem.  Section~\ref{antenna_selection_algorithms} presents the proposed antenna selection algorithms, and Section~\ref{evaluation_and_simulation} focuses on performance evaluation based on numerical simulations, leading to the conclusion in Section~\ref{conclusion}.

\section{Preliminaries and Problem Statement}
\label{problem_statement}

\subsection{Signal Model in Switch-Based Arrays}
We consider a switch-based uniform linear array (ULA) with $N$ antennas and $M$ RFCs {($M \ll N$)}. The antennas are uniformly spaced with an inter-antenna distance of $d$ ($\lambda/2$ units); hence, $d_m = (m-1)d$ is the distance between the first and $m$th antennas. Each of the $M$ RFCs can be connected to one of the $N$ antennas through a switch network. Considering all the possible $M$-antenna combinations, we can form a subarray set $\mathcal{B}_{N,M}$ with $F_{N,M} = {N \choose M}$ elements as
\begin{equation}
\mathcal{B}_{N,M} = \{\bv_{N,M}^1, ..., \bv_{N,M}^{F_{N,M}}\},
\label{antenna_selection_set}
\end{equation}
where each element $\bv_{N, M}^k$, representing a subarray formed by the selected antennas, is an $N\times 1$ selection vector with $M$ ones at the selected antenna indices and zero elsewhere.
The subarray set $\mathcal{B}_{N,M}$ contains all the possible antenna selection choices. 

The vector $\bv_{N, M}^k$ is used as the binary representation of the antenna selection, which is convenient for designing an antenna selection algorithm. To facilitate the problem formulation and lower bound derivative in the following section, we use another interchangeable representation for antenna selection, namely, an antenna location vector $\dv$ (in terms of half-wavelength). For instance, for $N=4, M=3$, and inter-antenna distance $d=1$, a selection example is given by
\begin{equation}
        \dv = 
            \begin{bmatrix} 
            0 & 1 & 3
            \end{bmatrix}^T, \,\, \text{or}\ 
        \bv =
            \begin{bmatrix} 
            1 & 1 & 0 & 1 
            \end{bmatrix}^T.
\end{equation}

{For a given subarray $\dv$, the single-snapshot observation of a narrowband signal $\yv\in\mathbb{C}^{M\times1}$ can be expressed as 
\begin{equation}
    \yv^{} = \alpha_0\av(u)^{}s^{} + \nv^{},
\label{DOA_signal_model}
\end{equation}}where $\alpha_0$ is the channel gain, {$s^{}$ is the transmitted signal symbol}, and $\nv^{} \in \mathbb{C}^{M\times 1}$ is the additive white Gaussian noise with a complex normal distribution $\mathcal{CN}(0, \sigma_n^2)$. We use $u=\sin(\theta) \in (-1, 1)$ to denote the signal coming from the direction $\theta$, and $\av(u)^{} = e^{-j\pi \dv^{} u} = [e^{-j\pi d_1u}, \ldots, e^{-j\pi d_{M}u}]^T \in \mathbb{C}^{M\times 1}$ is the steering vector. In this work, for simplicity, we consider the single-source/single-snapshot case. {Extension to multiple snapshots is straightforward, and extension to multiple sources is also possible~\cite{threshold_tsp}.}

\subsection{Maximum Likelihood DOA Estimation}
{To obtain a DOA estimate, algorithms such as MUSIC~\cite{doa_music} and ESPRIT~\cite{doa_esprit} can be used. However, the evaluation of DOA algorithms is beyond the scope of this work. We adopt a simple yet efficient maximum-likelihood estimator (MLE) to aid the antenna selection process, which works well in both the asymptotic and threshold regimes~\cite{doa_mle}.} Maximum-likelihood-based DOA estimation seeks to maximize the probability of receiving a signal vector
$\hat\yv$. The MLE is given by 
\begin{equation}
    \hat{u} = \argmax_u \av(u)^H \hat \Rm \av(u),
    \label{eq_beampattern}
\end{equation}
where $\hat \Rm = \hat\yv \hat\yv^H $ is an estimation of the received signal covariance matrix.

\subsection{The Antenna Selection Problem Statement}
For a signal received from a direction $u\in(-1, 1)$ with a signal-to-noise ratio (SNR) $S$, inter-antenna distance $d$ and a selection vector $\bv$, its DOA estimation MSE, using a certain algorithm, can be defined as $\mathrm{MSE}(u, S, d, \bv_{N, M})$. Our goal is to design an antenna selection algorithm to find the optimal $M$-antenna selection vector $\bv^*_{N,M} \in \mathcal{B}_{N,M}$ that minimizes expectation of DOA estimation MSE {(rather than the CRLB which does not work well in low SNR region)}. This problem can be formulated as
\begin{equation}
    \bv_{N, M}^* = \argmin_{\bv_{N, M} \in \mathcal{B}_{N,M}} \mathrm{MSE}(u, S, d, \bv_{N, M}).
\label{optimization_equation}
\end{equation}






\section{The Proposed Antenna Selection Algorithms}
\label{antenna_selection_algorithms}
In order to solve the problem formulated in~\eqref{optimization_equation}, two practical objective functions, the CRLB, and the PSL will be introduced in Section \ref{asof} before proceeding to the proposed TRA-based algorithms.

\subsection{{Benchmark Antenna Selection Objective Functions}}
\label{asof}

\subsubsection{Cram\'er-Rao Lower Bound (CRLB)}
The CRLB is a lower bound on the variance of unbiased estimators of a deterministic (fixed, though unknown) parameter~\cite{threshold_tsp}. For a single-source DOA estimation, the variance of an unbiased estimator $\hat u$ of $u$ is lower-bounded as~\cite{threshold_tsp,crlb}
\begin{equation}
    {\mathrm{MSE}{(\hat u)} \geq \mathrm{CRLB}(S, d, \bv_{N, M})} = \frac{1}{2\pi^2SU},
    \label{eq_CRLB}
\end{equation}
where the SNR $S$ and the array diversity $U$ are defined as
\begin{equation}
    S = \frac{MN_s |\alpha_0|^2}{\sigma_n^2}, \ U = \frac{1}{M}\sum_{i=1}^M \left(d_i-\frac{1}{M}\sum_{j=1}^M d_j\right)^2.
    \label{equation_snr}
\end{equation}

The CRLB will be used as a benchmark to evaluate DOA estimation performance. However, this bound is not practical in low SNRs as it grossly overestimates the achievable DOA performance in the \emph{threshold regions}~\cite{threshold_tsp}.

\subsubsection{Peak Sidelobe Level (PSL)}
Recent works in antenna selection utilize the information from the array beampattern by incorporating a peak sidelobe level constrains~\cite{roy2013sparsity,wang2014adaptive}.
Similar to the MLE objective function in~\eqref{eq_beampattern}, for a source at  $\text{DOA} = u_0$, the beampattern for direction $u$ is defined as~\cite{threshold_tsp}
\begin{equation}
    V(u_0, u) = \av^H(u)\Rm(u_0)\av(u) = \av^H(u)\av(u_0)\av^H(u_0)\av(u).
    \label{eq_beampattern-2}
\end{equation}

The mainlobe peak value is given by $V(u_0, u_0)$ and the value of the $k$-th sidelobe peak is $V(u_0, u_k)$ ($k\in(1,...,K)$), {where the value of $u_k$ depends on $u_0$, $d$ and $\bv_\mathrm{M}$, and  $K$ is the number of sidelobes}. The PSL can be expressed as~\cite{wang2014adaptive}
\begin{equation}
    \mathrm{PSL}(u_0, d, \bv_{N,M}) = \frac{\max{V(u_0, u_k)}}{V(u_0, u_0)}. 
    \label{eq_psl}
\end{equation}
Now, the formulation of the antenna selection problem by optimizing the CRLB in~\eqref{eq_CRLB} with a constraint on PSL (PSL-C) in~\eqref{eq_psl} can be expressed as
\begin{equation}
    \begin{split}
        & \bv^*_{N,M} = \argmin_{\bv_{N, M} \in \mathcal{B}_{N,M}} \mathrm{CRLB}(S, d, \bv_{N, M})\\
        & \text{s.t. } \mathrm{PSL}(u_0, d, \bv_{N, M}) \le \delta, \delta\in [0,1],
    \end{split}
    \label{eq_crlb_psl}
\end{equation}
where $\delta$ is the constrain threshold. This problem can be solved by using an exhaustive search or via convex relaxation methods, as proposed in~\cite{wang2014adaptive}. The main drawback of this method is the requirement to tune the threshold parameter $\delta$, which may vary with SNR, subarray layout, and source direction. {In the following section, we describe the approximation of DOA estimation MSE in the threshold region and the proposed TRA-based antenna selection algorithms.}

\subsection{MSE Threshold Region Approximation}
To approximate expected DOA performance in terms of MSE, we analyze the peaks of the noise-free beampattern $V(u_0, u)$ for a specific source location $u_0$~\cite{threshold_tsp}. With a properly designed array layout, $V(u_0, u_0)$ should be the {highest} peak of the beampattern.
In the presence of noise, the signal covariance matrix $\hat\Rm$ (in~\eqref{eq_beampattern}) is different from $\Rm(u_0)$ in~\eqref{eq_beampattern-2}. The location of the main peak may change, which causes errors in DOA estimation {due to the occurrence of \emph{outliers}}. 
By calculating the probability of an outlier $P_k$ occurring at sidelobe peak $u_k$ and the probability of non-outlier $1-\sum_{k=1}^K P_k$, the approximated MSE can be expressed as~\cite{threshold_expression}
\begin{equation}
\begin{split}
    & \mathrm{TRA}(u_0, S, d, \bv_{N, M})\\
    & \approx \left(1-\sum_{k=1}^K P_k\right)\mathrm{CRLB}(S,d,\bv_{N,M}) + \sum_{k=1}^K P_k(u_k-u_0)^2,
\end{split}
\label{eq_expectation}
\end{equation}
where $K$ is the number of sidelobes and 
\begin{equation}
    P_k \approx \frac{1}{2}\exp(\frac{-S}{2})I_0(\frac{S|\av^H(u_0)\av(u_n)|}{2M}),
    \label{eq:probability_sidelobe}
\end{equation}
with $I_0(\cdot)$ being the modified Bessel function of the first kind and order zero. {The expression in~\eqref{eq_expectation} is obtained by approximating the DOA estimation error as the statistical average over all the possible errors with the estimate occurring on the mainlobe (CRLB component) and errors that move the estimation to any of the sidelobes (peak $1$ to peak $K$).} {For more details on~\eqref{eq_expectation}, the readers are referred to~\cite{threshold_tsp, threshold_expression}}

Evaluating the MSE in \eqref{eq_expectation} requires a prior DOA information $\hat u$ (denoted as $u_0$ in~ \eqref{eq_expectation}) and the SNR $S$ as inputs.
{We assume that $S$ is known (e.g., by using techniques in~\cite{pauluzzi2000comparison}), and $\hat u$ can be obtained via extra sensors such as inertial measurement units (IMUs) and GPS, or estimated with the current selected subarray, which may not be accurate.}

To improve the robustness of the selection algorithm to inaccurate DOA estimation, we utilize $N_A$ anchor angles. Assume that $\hat u \in [u-\Delta u, u+\Delta u]$ is the initial DOA estimation, where $\Delta u$ is the maximum estimation error. Depending on the estimation accuracy of the initial DOA and computational limitations of the antenna selection algorithm, a set of $N_A$-anchors can be selected as $\uv_{\text{anchor}} = \{u_1, u_2, \cdots, u_{N_A}\}$, where $u_n \in [\hat u-\Delta u, \hat u + \Delta u]$.
Now, the TRA-based antenna selection problem can be formulated by minimizing the worst-case MSE approximation as 
\begin{equation}
    \bv^*_{N,M} = \argmin_{\bv_{N,M} \in \mathcal{B}_{N,M}} \left(\argmax_{u_{n} \in \uv_{\text{anchor}}} \mathrm{TRA}(u_n, S, d, \bv_{N,M})\right).
\label{TRA_minmax}
\end{equation}
Note that performing the optimization~\eqref{TRA_minmax} could be computationally demanding because of the large number of elements in $\mathcal{B}_{N,M}$.
In Sections~\ref{sec:the_proposed_algorithms}, we first introduce the subarray layout alignment to remove the redundant subarrays. Further, we develop two low-complexity algorithms to solve the optimization problem in~\eqref{TRA_minmax}.




\subsection{The Proposed Antenna Selection Algorithms}
\label{sec:the_proposed_algorithms}

{Due to the fact that some subarrays have identical beampatterns, they provide the same DOA performance and thus are redundant. This section first describes the subarray layout alignment process to produce a unique subarray set, based on which two antenna selection algorithms are proposed.}

\begin{figure}[t]
\centering
\includegraphics[width=2.3 in]{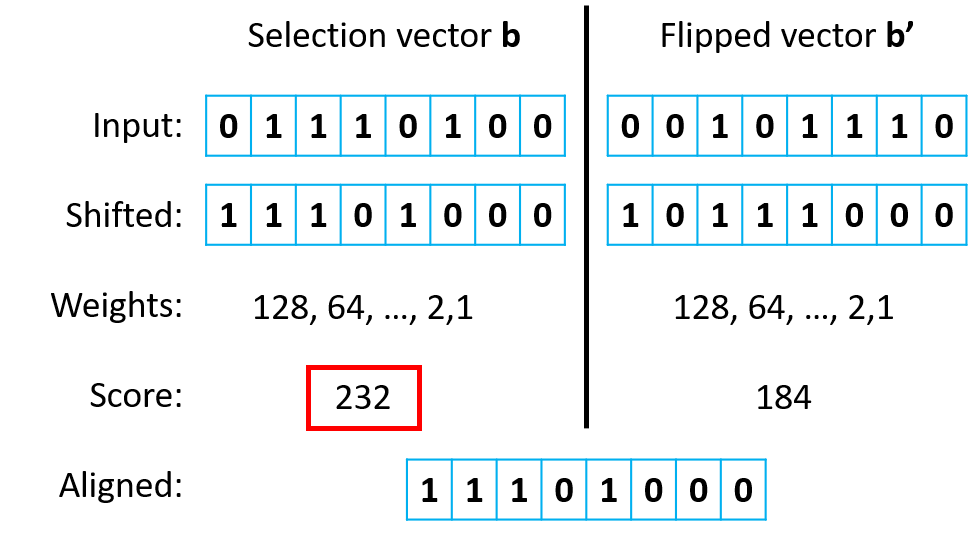}
\caption{Illustration of the subarray layout alignment process.}
\label{fig-alignment_illustration}
\end{figure}

\begin{algorithm}[t]
\footnotesize
\caption{Subarray Layout Alignment}
\label{alg_alignment}
\begin{algorithmic}[1]
\State Input: $\bv$
\While{\bv(1) == 0}
    \State $\bv \leftarrow \mathrm{leftShift}(\bv)$
\EndWhile
\State $\bv' \leftarrow \mathrm{flip}(\bv)$
\While{$\bv'$(1) == 0}
    \State $\bv' \leftarrow \mathrm{leftShift}(\bv')$
\EndWhile
\State $\bv_{\text{aligned}} = \argmax_{\{\bv,\bv'\}}(\mathrm{score}(\bv),\mathrm{score}(\bv'))$
\State
\Return $\bv_{\text{aligned}}$
\end{algorithmic}
\end{algorithm}

\subsubsection{Subarray Layout Alignment}
\label{sec:layout_alignment}

In order to merge redundant subarrays in the set~$\mathcal{B}_{N,M}$, a subarray layout alignment process is performed before delving into the antenna selection process. To illustrate subarray layout alignment, we take the case where $N=8$ and $M=4$ for example. In this case, subsets that are {circularly shifted versions of each other}, such as $\bv_{8,4}^1 = [0,0,0,1,1,1,0,1]^T$ and $\bv_{8,4}^2 = [0,1,0,1,1,1,0,0]^T$, are equivalent. The purpose of this subarray layout alignment is threefold: a) to reduce the search space represented by the antenna selection set; b) to reduce the number of switches in the circuit design; and c) to improve the convergence of the {neural network (NN) training process for the TRA-DL algorithm {(see Section~\ref{sec_tra_dl} further ahead)}}.

The pseudocode for the subarray layout alignment process is listed in Algorithm~\ref{alg_alignment}.
For a selected subarray $\bv$, the layout alignment process first circularly left-shifts $\bv$ till the first element is `$1$' to obtain $\bv_s$. Then, we want the `\textit{mass center}' of the layout to be on the left and the $n$-th antenna will be assigned a weight $2^{n-1}$. The score of the shifted vector $\bv_s$ is calculated as
\begin{equation}
    \mathrm{score}(\bv) = \sum_{n=1}^N 2^{b-1}\bv_s(n). 
\end{equation}
The same procedure is repeated to the flipped version of $\bv$ denoted as $\bv'$ and the final aligned antenna set $\bv_{\text{aligned}}$ is chosen from the shifted vectors $\bv$ and $\bv'_s$ with the higher score. One example of subarray layout alignment is shown in Fig.~\ref{fig-alignment_illustration}.

By performing subarray layout alignment, we obtain a non-redundant set $\tilde{\mathcal{B}}_{N,M}$. 
We define  $F_{N,M}$, $\tilde F_{N,M}$, $G_{N,M}$, $S_{N,M}$ and $\tilde S_{N,M}$ as the number of full subarray set (cardinality of $\mathcal{B}_{N,M}$), number of unique subarray set (cardinality of $\tilde{\mathcal{B}}_{N,M}$), number of subarrays using greedy search, number of switches needed for fully connected architecture, and number of switches needed for the unique subarray set, respectively. The parameters for different $N$ and $M$ values are shown in Table~\ref{table:unique_set}, with the ratio $U_{N,M}/F_{N,M}$ also given. We can see that the number of unique subarrays is much smaller than the total number of subarrays\footnote{{Although the number of unique subarrays is reduced, the computational cost for a large $N$ is still high, which needs to be considered in array design.}}.

\begin{table}[t]
\footnotesize
\caption{Number of Redundant and Unique Sets}
\label{table:unique_set}
\centering
\begin{tabular}{c|c c c | c | c c}
\hline
\textbf{$N$, $M$} &  $F_{N,M}$ & {$\tilde F_{N,M}$} & \text{Ratio} & $G_{N, M}$ & $S_{N,M}$ & $\tilde S_{N, M}$\\
\hline
11, 2  & 55 & 10 & 0.1818 & 63 & 22 & 11\\
11, 4  & 330 & 70 & 0.2121  & 56 & 44 & 21\\
11, 6  & 462 & 136 & 0.2944 & 45 & 66 & 27\\
21, 4  & 5985 & 615 & 0.1028 & 221 & 84 & 46\\
21, 6  & 54264 & 7872 & 0.1451 & 210 & 126 & 72\\
21, 8  & 203490 & 38970 & 0.1915 & 195 & 168 & 90\\
\hline
\end{tabular}
\end{table}

\subsubsection{A Greedy Approach based on TRA (TRA-G)}
It is not feasible to use a brute-force approach to find the optimal subarray, which requires evaluating the objective function~\eqref{eq_expectation} for all elements of $\tilde{\mathcal{B}}_{N, M}$. As a practical solution, we propose a \emph{greedy} approach based on the threshold region approximation (TRA-G) to find the optimal antenna set that minimizes DOA estimation MSE in an efficient manner.
To develop our greedy algorithm, we start with the following {\bf{assumption}}: The optimal antenna selection vector $\bv^*_{N,M-1}$ of the $(m-1)$-RFC case shares the same `$1$' elements with the optimal antenna selection vector $\bv^*_{N,m}$ of the $m$-RFC case ($2 \le m \le N$), which can be expressed as ${\bv^*_{N,m-1}}^T{\bv^*_{N, m}} = m-1$. Although this approach is not guaranteed to return a globally optimal solution, as we will see later, it provides satisfactory results.

The pseudocode for the greedy algorithm TRA-G is listed in Algorithm~\ref{alg_TRA_Greedy}.
To obtain $\bv^*_{N,M}$, we need to firstly search for $\bv^*_{N,N-1} \in \tilde{\mathcal{B}}_{N, N-1}$ with the minimum DOA MSE approximation. Then, we form another collection of sets $\tilde{\mathcal{B}}_{N-1, N-2}$ from $\bv^*_{N-1}$ and search for $\bv^*_{N,N-2}$, and continue until we find $\bv_{N,M}^*$. 
In this way, $G_{N,M}={(N+M+1)(N-M)}/{2}$ calculations of $\mathrm{TRA}(\uv_\text{\text{anchor}}, S, d, \bv_{N,M})$ are needed\footnote{Note that the number of calculations is not always reduced by using the greedy method, especially when $M$ is small, e.g., $\tilde F_{11,2}$=$10<$ $G_{11,2}$=$63$. Thus the exhaustive search can be adopted when necessary.}.

\begin{algorithm}[t]
\footnotesize
\caption{TRA-G}
\label{alg_TRA_Greedy}
\begin{algorithmic}[1]
\State Input: $\uv_\text{\text{anchor}}, S, d, N, M$
\State $\bv_{N,N}^* \leftarrow \bv_{N,N}$
\For{$j = 0$ to $(N-M-1)$}
    \State Create $\mathcal{B}_{N-i, N-i-1} = \{\bv^1_{N-i-1},...,\bv^{N-i}_{N-i-1}\}$ from $\bv^*_{N, N-i}$
    \For{$i = 1$ to $N-i$}
        \State Calculate $\mathrm{CRLB}(S, d, \bv_{N,N-i-1}^j)$~\eqref{eq_CRLB}
        \State Calculate $V(u_n, u)$ for $u_a\in\uv_\text{\text{anchor}}$~\eqref{eq_beampattern-2}
        \State Calculate $\mathrm{TRA}(u_n, S, d, \bv_{N,N-i-1}^j)$ for $u_a\in\uv_\text{\text{anchor}}$~\eqref{eq_expectation}
    \EndFor
    \State $\bv^*_{N, N-i-1}$
    \State $\ \ \ \ = \argmin_j (\argmax_{u_n} \mathrm{TRA}(u_n, S, d, \bv_{N,N-i-1}^j)))$~\eqref{TRA_minmax}
\EndFor
\State
\Return $\bv^*_{N,M}$
\end{algorithmic}
\end{algorithm}

\begin{algorithm}[t]
\footnotesize
\caption{{TRA-DL}}
\label{alg_TRA_DL}
\begin{algorithmic}[1]
\State \textbf{---\textit{Training Phase}---}
\State Input: Training Data Set $\mathcal{D}$
\While{Error Not Converge}
\State Select a fraction of the training Data Set $\mathcal{\tilde D}\in \mathcal{D}$
\For{$i = 1$ to $|\mathcal{\tilde D}|$}
    \State $\langle \xv_{0,i}, \hat \bv^*_{N,M,i}\rangle$ $\leftarrow$ $D_{i}$
    \State $\hat b_i = f_\text{NN}(\wv, \xv_{0,i})$~\eqref{eq:nn_forward},\eqref{eq:nn_layer}
    \State Calculate $\frac{\partial \mathcal{L}_i}{\wv}$
\EndFor
\State Update parameters $\wv$ based on the selected optimizer
\EndWhile
\State
\Return $\wv^*$
\State \textbf{---\textit{Testing Phase}---}
\State Input: $\hat x = \langle u, S\rangle$
\State $\hat \bv = f_\text{NN}(\wv^*, \hat \xv)$~\eqref{eq:nn_forward},\eqref{eq:nn_layer}
\State
\Return Indices of top-$M$ elements in $\hat \bv$
\end{algorithmic}
\end{algorithm}

\subsubsection{A Deep Learning Approach based on TRA (TRA-DL)}
\label{sec_tra_dl}
{For a real-time system, especially when {$N \choose M$} is large, the computational cost of the TRA-G algorithm might be high.} In this case, a pre-trained NN could be used to approximate the TRA-G algorithm. {We use the greedy method to generate the training dataset $\mathcal{D}$ for the NN with each element containing a prior-subarray pair as $\mathcal{D}_i = \langle \xv_{0,i}, \bv^*_{N,M,i} \rangle$ ($\xv_{0,i} = [u_i, S_i]^T$). Note that the NN is trained for a specific array setup (e.g., a fixed number of antennas and their spacings). Any array layout changes require the model to be retrained.} 

{A multi-layer perceptron (MLP) NN is used for this task with $H$ layers and $g_h$ ($1\le h \le H$) elements in each layer. We define $g_0=2$ and $g_{H+1} = N$ as the number the elements of the input layer (DOA prior and SNR) and output layer (score of each antenna), respectively. The output of the $h$th layer $\xv_h$ ($1\le x\le H+1$) can be obtained by using a weight matrix $\Wm_g\in \mathbb{R}^{g_{h-1}\times g_{h}}$, a bias vector $\wv_g\in \mathbb{R}^{g_{h}\times 1}$, and an activation function (e.g., a rectified linear unit (ReLU)) as
\begin{equation}
    \xv_{h} \triangleq g_{\text{acti},h}(\Wm_{h}, \wv_h, x_{h-1}) =
    \text{ReLU}(\xv_{h-1}\Wm_h + \wv_h).
    \label{eq:nn_layer}
\end{equation}
The output of the NN can then be obtained as
\begin{align}
    \hat \bv \triangleq f_\text{NN}(\wv,u, S) =\text{sigmoid}( \xv_{h+1}),
    \label{eq:nn_forward}
\end{align}
where $\wv$ is a vector containing all the $\sum_{h=1}^{H+1} (g_{h-1} g_{h} + g_{h})$ network parameters, and $\text{sigmoid}(x) = 1/(1+e^{-x})$ is the sigmoid function.}

We use the following mean squared error loss as the objective function of the NN:
\begin{equation}
    \mathcal{L} = \frac{1}{N}\sum(\hat\bv-\bv^*_{N,M})^2,
    \label{dl_loss_function}
\end{equation}
where $\bv^*_{N,M}$ is the optimal antenna set obtained using the TRA-G algorithm. The purpose of the training phase is to obtain the optimal network parameters $\wv^*$ such that the loss in~\eqref{dl_loss_function} is minimized for all the training dataset $\mathcal{D}$. 
Finally, the trained model can be used to perform antenna selection by taking the prior (or estimated) source direction and SNR $\hat \xv_0 = [u, S]^T$ as the input, and the top-$M$ antennas with the highest score in the vector $\hat \bv = f_\text{NN}(\wv^*, \xv_0)$ can be chosen as the selected subset.
The pseudocode for the DL-based algorithm TRA-DL is listed in Algorithm~\ref{alg_TRA_DL}. 

\section{Performance Evaluation}
\label{evaluation_and_simulation}

\subsection{Simulation Setup}
\label{simulation_setup}
In our simulations, we test a scenario with a receiver uniform linear array (ULA) of size $N = 21$, inter-antenna distance $d = 0.5$ ($\equiv\lambda/4$), and a {uniformly distributed} source direction $u\in [-0.9, 0.9]$. 
For the neural-network-based sensor selection algorithm (\mbox{TRA-DL}), we choose $\gv = [16, 32, 64, 32, 16]$ elements for each layer. An Adam optimizer~\cite{kingma2014adam} with parameters ($lr = 0.001, \beta_1 = 0.9, \beta_2 = 0.999, \epsilon = 1e-7$, number of iterations as $200$) and mean squared error loss function in~\eqref{dl_loss_function} are used for training. We randomly generate 10$K$ samples within specific SNR and DOA ranges to form a training dataset $\mathcal{D}$.

The selected subarrays using \mbox{TRA-G} and \mbox{TRA-DL} are evaluated and compared with a number of benchmarks; namely, the $M$-antenna ULA with a $\lambda/2$ spacing, the constrained PSL {PSL-C}, and the best case CRLB among all possible subarrays in $\tilde{\mathcal{B}}_{N,M}$.
After antenna selection, an MLE with a coarse search of $0.2^\circ$ and a fine search of $0.01^\circ$ is performed
to obtain the final DOA estimation. 
The averaged MSE computed from simulation is used as a performance indicator for different selection algorithms. Matlab code is available at~\cite{chenhui07c8}.

\begin{figure*}[!ht]
\begin{minipage}[b]{0.315\linewidth}
  \centering
  \centerline{\includegraphics[width=0.95\linewidth]{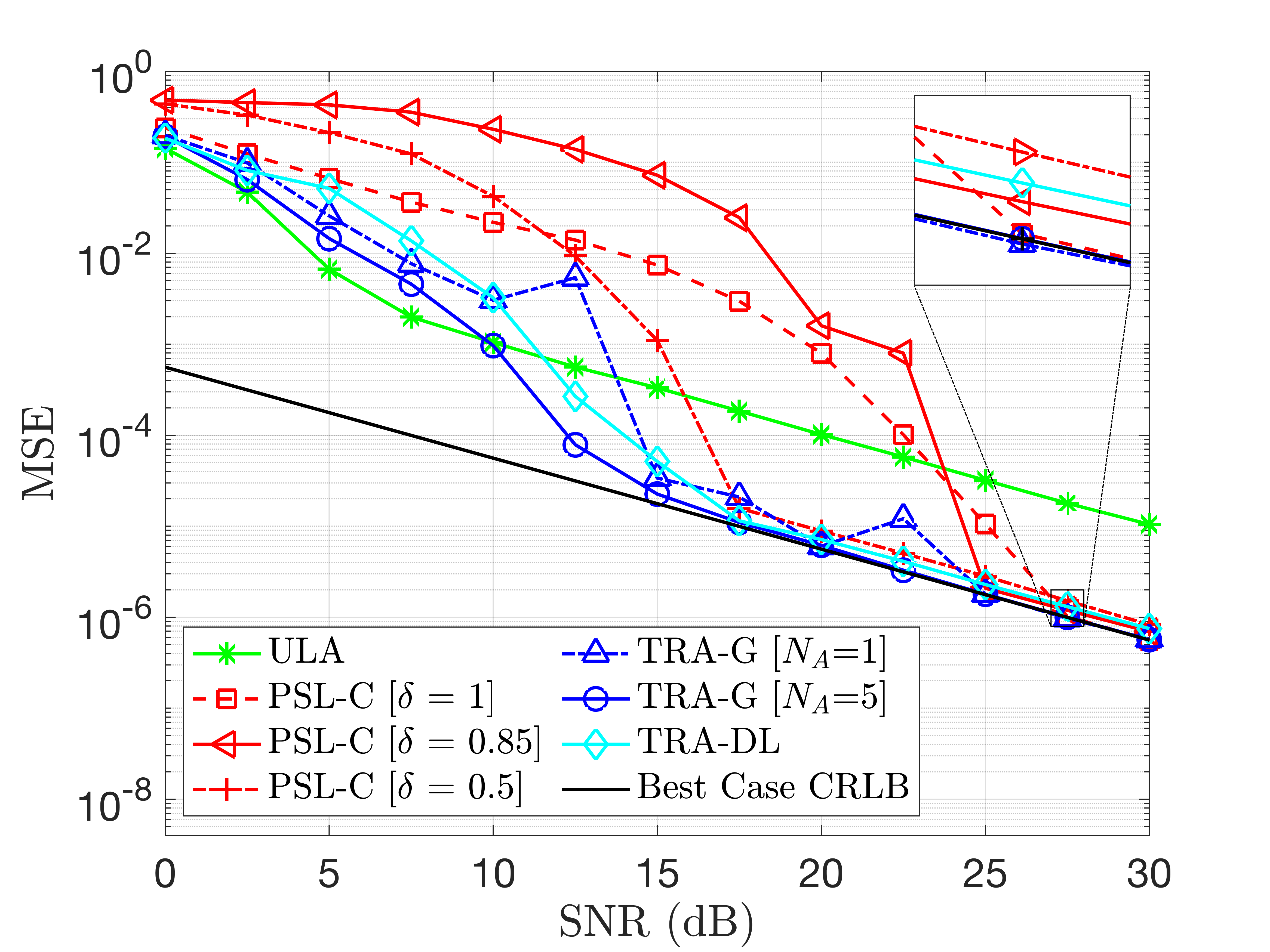}}
  \centerline{(a) $M = 4$} \medskip
\end{minipage}
\hfill
\begin{minipage}[b]{0.315\linewidth}
  \centering
  \centerline{\includegraphics[width=0.95\linewidth]{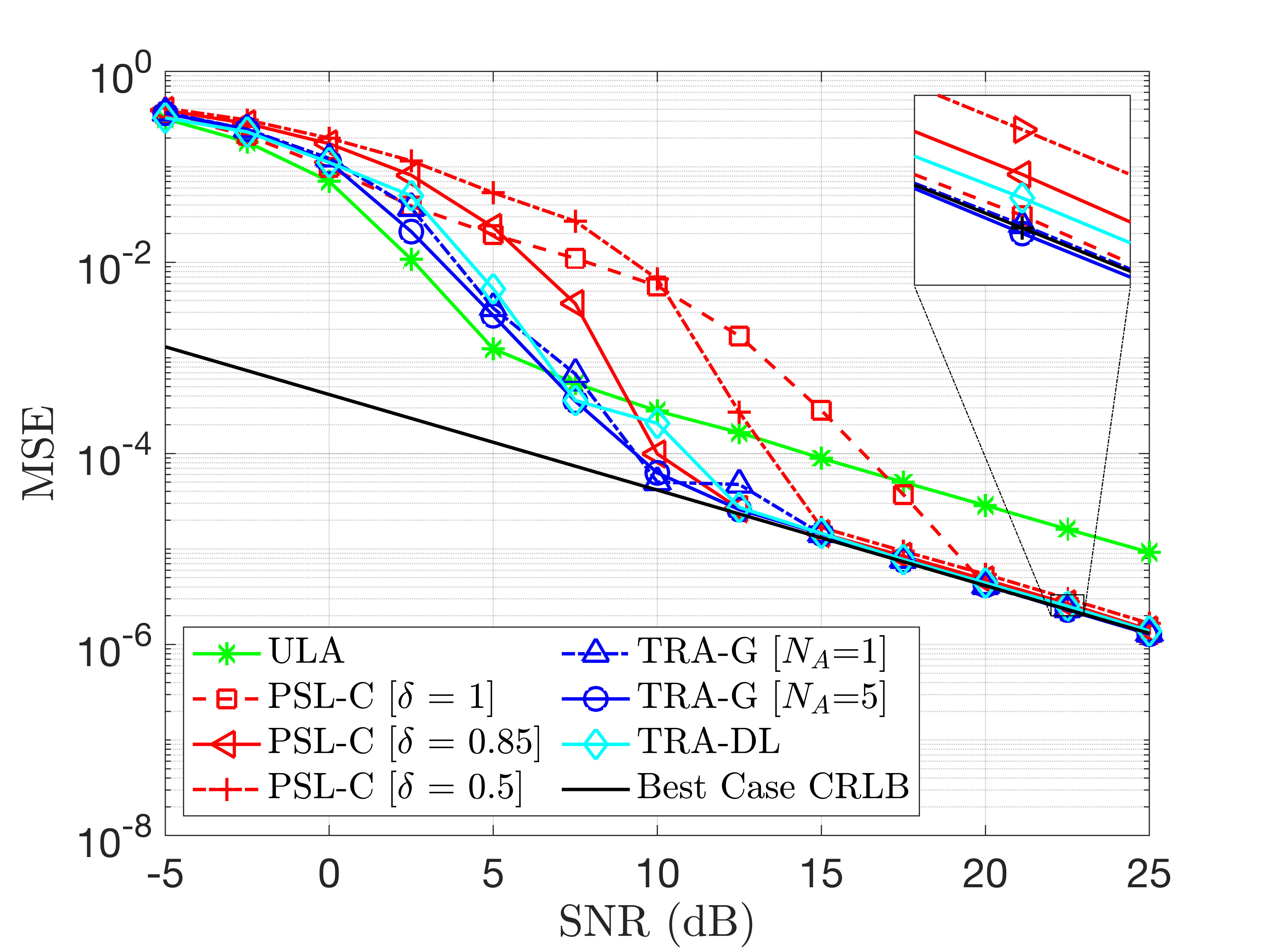}}
  \centerline{(b) $M = 6$} \medskip
\end{minipage}
\hfill
\begin{minipage}[b]{0.315\linewidth}
  \centering
  \centerline{\includegraphics[width=0.95\linewidth]{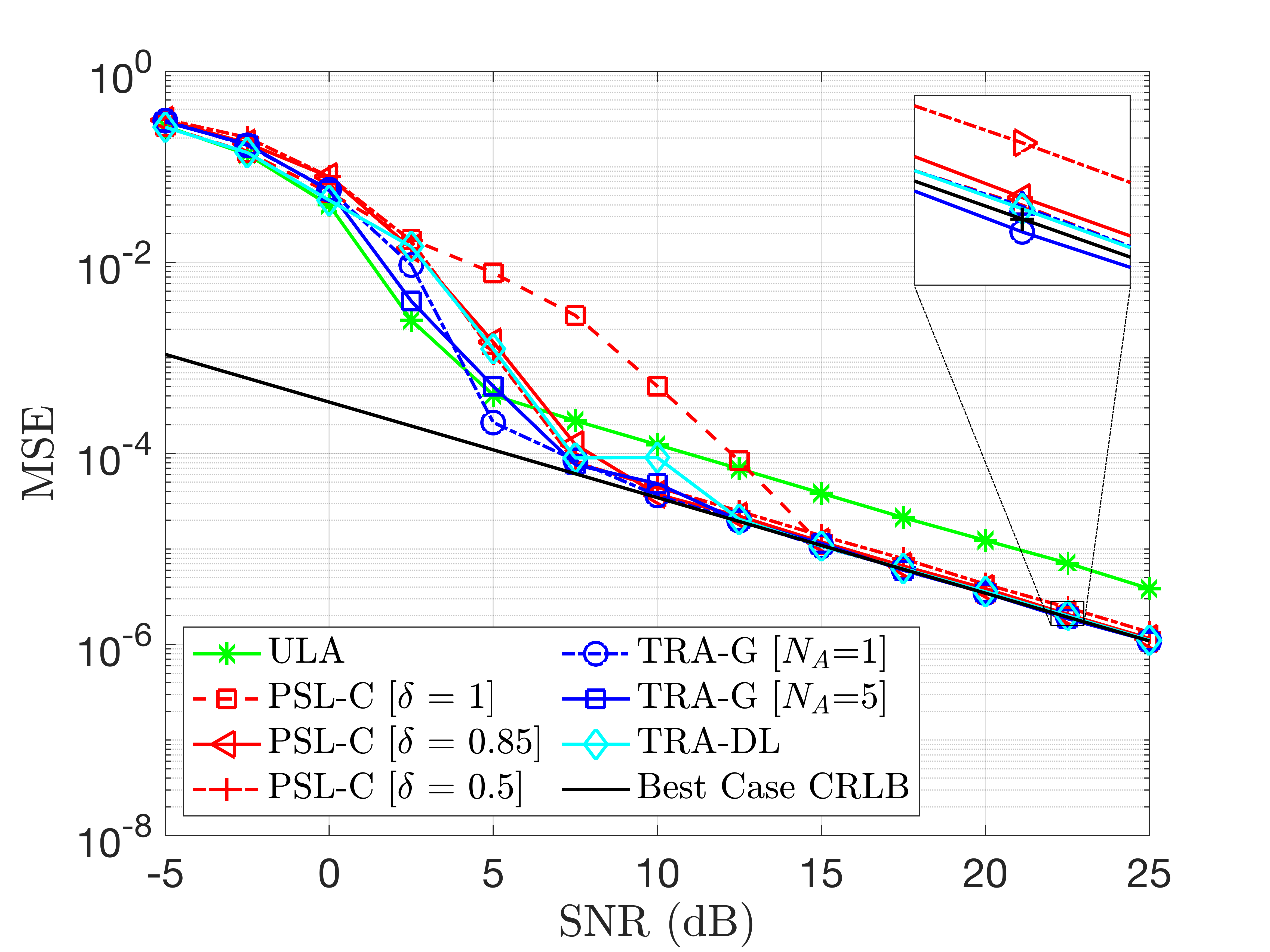}}
  \centerline{(c) $M = 8$} \medskip
\end{minipage}
\hfill
\vspace*{-2 mm}
\caption{DOA simulation results using different antenna selection algorithms: (a) $M=4$, (b) $M=6$, (c) $M=8$. {The total number of antennas is $N=21$.}}
\label{fig-different_M}
\end{figure*}

\subsection{{Single-Snapshot DOA Estimation}}
\label{simulation_with_prior}
The performance of the two proposed algorithms, TRA-G and TRA-DL, is evaluated. PSL-C in~\eqref{eq_crlb_psl} is used as a benchmark algorithm with different values of the constraint threshold $\delta = \{1, 0.85, 0.5\}$. Note that $\delta=1$ indicates no PSL constraint and the optimization relies soley on the CRLB (this yields $\dv_{\delta=1} = [0, 0.5, 1, 9, 9.5, 10]$ for $M=6$). In addition, a ULA setup (e.g., {a linear array with} $\dv_\text{ula} = [0,1,2,3,4,5]$ for $M=6$) and a best-scenario CRLB (e.g., the CRLB of $\dv_{\delta=1}$ for $M=6$) are also used as benchmarks. {We assume a coarse DOA estimate is available (e.g., from current connected subarray, or from IMUs and GPS), and the maximum DOA estimation error $\Delta u$ is chosen as $0.1$ for generating $\hat u$.} The number of anchors $N_A$ is chosen as $1$ and $5$ ($u_a = \hat u+\frac{a-3}{2}\Delta u$, $a=1,2,\cdots, 5$). The {single-snapshot DOA estimation MSE} is calculated from $5000$ simulation trials at each SNR point. Results for $M=4, 6, 8$ are shown in Fig.~\ref{fig-different_M}.

{From Fig.~\ref{fig-different_M}, we can see that the proposed greedy antenna selection method TRA-G improves DOA estimation performance, especially when $M$ is small, e.g., $M=4$ as shown in~Fig.~\ref{fig-different_M} (a).
Although the TRA-DL algorithm performs not as well as the greedy algorithms, it provides a fast option for antenna selection.
The PSL-C algorithm works well when $M$ is large and reaches the CRLB when the SNR is high. However, the selection of the constraint threshold $\delta$ affects its performance.
The ULA layout works well in the low SNR regime; this is due to the approximation error of the MSE close to the no-information region. As evident from the figure, the proposed antenna selection algorithms yield the best performance without perfect knowledge of the true DOA}.

\begin{figure}[t]
\centering
\begin{minipage}[b]{0.9\linewidth}
  \centerline{\includegraphics[width=0.99\linewidth]{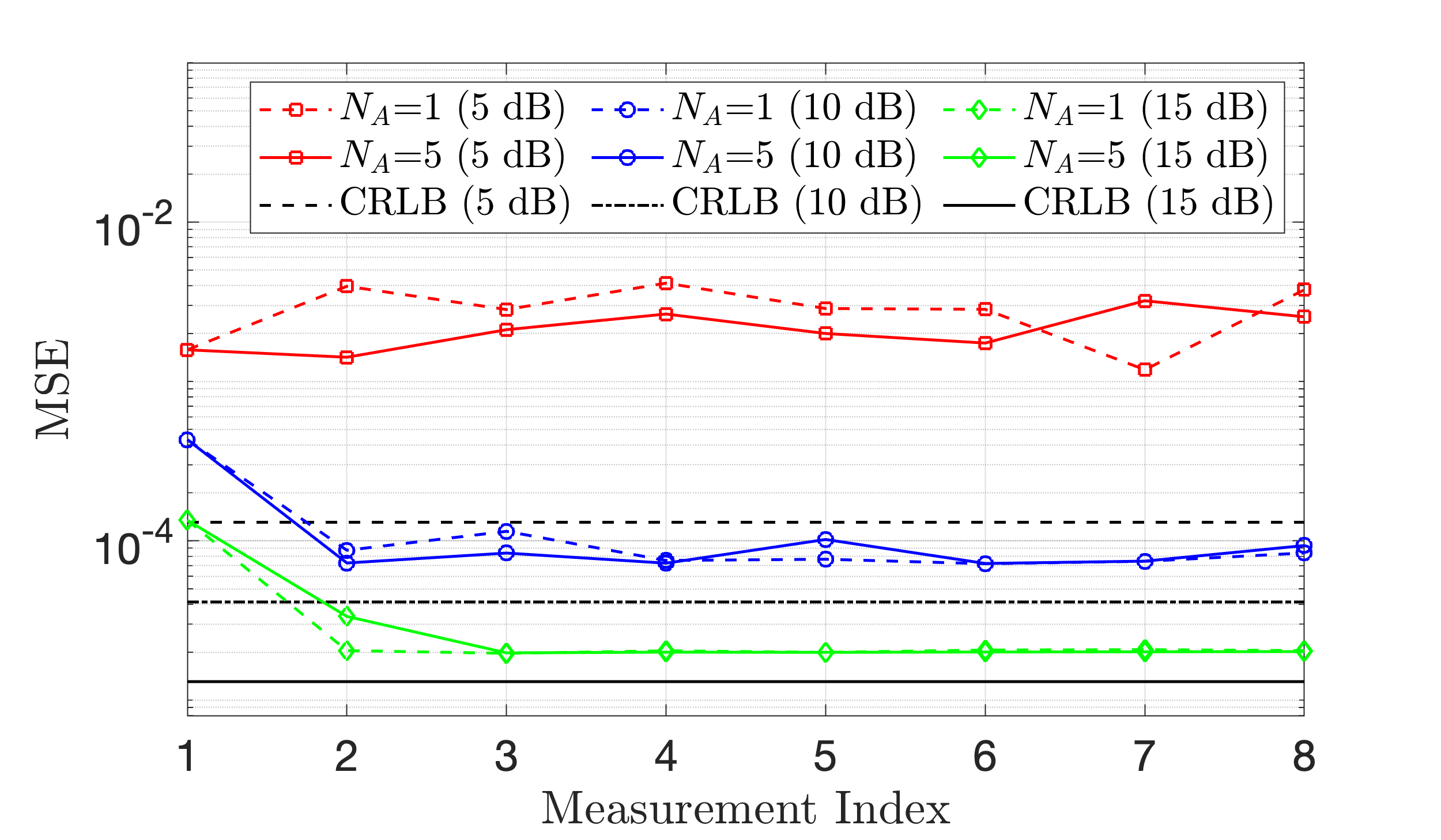}}
\end{minipage}
\vspace*{-2 mm}
\caption{DOA estimation with sequential measurements. The simulation parameters are the same as in Fig.~\ref{fig-different_M} (b), with the SNR equals to $\unit[5/10/15]{dB}$.}
\label{fig-different_snapshot}
\end{figure}

{
\subsection{Single-Snapshot DOA Estimation with Sequential Measurements}
With multiple measurements, DOA estimation and antenna selection can be performed sequentially. We can start with a default ULA subarray choice to obtain a DOA estimate using the first measurement. The estimated results are then used to select antennas for the subsequent measurement and obtain a new DOA estimate. The process carries on iteratively whereby a subarray is selected and a DOA estimate is calculated in each measurement. By using the same simulation parameters as in Fig.~\ref{fig-different_M} (b), the DOA estimation results (for SNR equals $5$ dB, $10$ dB, and $15$ dB) versus measurement number\footnote{The CRLB is calculated for each measurement. However, the CRLB considering all the observed signals from multiple measurements can also be derived, which is not discussed here.} are shown in Fig.~\ref{fig-different_snapshot}. We can see gaps between the simulation results and the CRLBs due to inaccurate DOA estimation using a default subarray. However, we can see that the antenna selection algorithms converge in $1$-$2$ snapshots. 
In the threshold region ($\unit[10]{dB}$ SNR), the results are less stable compared with the results in the high SNR region ($\unit[15]{dB}$ SNR). When the SNR is low, the antenna selection algorithms yield limited improvements.}

\subsection{Computational Complexity}
{We use the number of multiplication operations to indicate the complexity of different antenna selection algorithms as shown in Table~\ref{table:complexity}. For simplicity, constant multiplication operations that do not depend on the variables are ignored. The PSL-C algorithm requires $6N$ multiplications to define the array position matrix and ambiguity matrix, and $MF_{N,M} + 4M \check F_{N,M}$ multiplications for antenna selection, where $F_{N,M} = {N\choose M}$ can be found in Table~\ref{table:unique_set}, and $\check F_{N,M}$ is the number of antenna sets that satisfy the constraint. The TRA-G algorithm needs $G_{N,M} = {(N+M+1)(N-M)}/{2}$ executions of TRA($\cdot$) in~\eqref{eq_expectation}, which consists of finding the lobe peaks ($2NN_R$), calculating the CRLB ($N$), and obtaining the probability $P_k$ for $K$ sidelobes ($2N+2+\delta_B$), per~\eqref{eq_expectation} and~\eqref{eq:probability_sidelobe}. Here, $N_R$ is the number of DOA grid points used to calculate the number of sidelobes, and $\delta_B$ is the multiplication operations needed to calculate the modified Bessel function.
For a well-trained model, the TRA-DL only needs a fixed number of $\sum_{h=1}^{H+1}g_{h-1} g_{h}$ ($5488$ in this simulation) multiplications, which depends on the network structure, regardless of the value of $M$. Although the TRA-DL algorithm cannot outperform TRA-G, it offers a significant complexity reduction.}

\begin{table}[!hbt]
\caption{{Multiplication Operations Needed for Different Algorithms}}
\renewcommand{\arraystretch}{1.25}
\label{table:complexity}
\centering
\begin{tabular}{c|c }
\hline
\textbf{Method} & 
{\textbf{Number of Multiplication Operations}}\\
\hline
PSL-C  & $6N + MF_{N,M} + 4M \check F_{N,M}$ \\
TRA-G  & $G_{N,M} (2N N_R + N + K(2N+2+\delta_B))$ \\
TRA-DL & $\sum_{h=1}^{H+1}g_{h-1} g_{h}$ \\
\hline
\end{tabular}
\renewcommand{\arraystretch}{1}
\end{table}

\section{Conclusion}
\label{conclusion}
{We considered the antenna selection problem to improve DOA estimation performance in switched antenna array systems. We first perform a subarray layout alignment process to create a unique subarray set. Based on a threshold region approximation and the created unique subarray set, we proposed a greedy (TRA-G) and a deep-learning-based (TRA-DL) antenna selection algorithms.} The deep-learning-based algorithm is trained based on the results of the greedy algorithm. Numerical results show that our proposed greedy-based TRA-G algorithm provides superior performance compared to other benchmark algorithms, while the proposed deep learning-based TRA-DL algorithm reduces the computational complexity at the expense of a slight DOA estimation performance degradation.
Possible future directions include designing algorithms for multipath and 2-D array scenarios, in addition to exploiting other hybrid array architectures and neural network structures.


\ifCLASSOPTIONcaptionsoff
\fi
\bibliographystyle{IEEEtran}
\bibliography{reference}

\end{document}